\documentclass[epj,nopacs]{svjour}
\usepackage{epsfig,amsmath}
\usepackage{graphicx}

\newcommand{\nn}{\nonumber}

\newcommand{\Mpl}{\overline{M}_{\rm Pl}}

\newcommand{\gld}{\tilde G}
\newcommand{\go}{\tilde g}
\newcommand{\sq}{\tilde q}

\begin{document}

\title{HELAS and MadGraph with goldstinos}

\author{
 K.~Mawatari\inst{1}\fnmsep\thanks{e-mail: 
            kentarou.mawatari@vub.ac.be}
 \and 
 Y.~Takaesu\inst{2}\fnmsep\thanks{e-mail: takaesu@post.kek.jp} }
\institute{
 Theoretische Natuurkunde and IIHE/ELEM, Vrije Universiteit Brussel,\\
 and International Solvay Institutes,
 Pleinlaan 2, B-1050 Brussels, Belgium
 \and
 KEK Theory Center, and Sokendai, Tsukuba 305-0801, Japan
 }

\abstract{
Fortran subroutines to calculate helicity amplitudes with
goldstinos, which appear as the longitudinal modes of massive gravitinos
in high energy processes, are added to the {\tt HELAS} 
({\tt HEL}icity {\tt A}mplitude {\tt S}ubroutines) library.
They are coded in such a way that arbitrary amplitudes with external
goldstinos can be generated automatically by {\tt MadGraph}, after slight
modifications. All the codes have been tested carefully by making use of 
the goldstino equivalence theorem and the gauge invariance of the
helicity amplitudes. 
Hadronic total cross sections for associated 
gravitino productions with a gluino and a squark are also presented. 
}

\titlerunning{HELAS and MadGraph with goldstinos}
\authorrunning{K.~Mawatari and Y.~Takaesu}

\maketitle

\vspace*{-100mm}
\noindent KEK-TH-1437 
\vspace*{83mm}


\section{Introduction}\label{intro}

Goldstinos are Goldstone fermions, massless spin-1/2 particles,
associated with spontaneous supersymmetry (SUSY) breaking,
and appear as the helicity $\pm1/2$ states of massive gravitinos via the
super-Higgs mechanism in 
local supersymmetric extensions to the Standard Model (SM).
While the interactions of the helicity $\pm3/2$ components of the
gravitino are suppressed by the Planck scale, those of the helicity
$\pm1/2$ components are suppressed by the SUSY breaking scale and 
can be important even for collider phenomenology 
in low-scale SUSY breaking scenarios, e.g., gauge-mediated
SUSY breaking~\cite{Giudice:1998bp}.

In the recent paper~\cite{Hagiwara:2010pi} K.~Hagiwara and the authors
introduced new {\tt HELAS} subroutines~\cite{Hagiwara:1990dw} to
calculate helicity amplitudes with massive spin-3/2 gravitinos.
They are coded in such a way that
arbitrary amplitudes with external gravitinos can be generated
automatically by {\tt MadGraph} \cite{Stelzer:1994ta}.

In this paper, taking into account high energy processes with the
center-of-mass (CM) energy $\sqrt{s}\gg m_{3/2}$,
we present new {\tt HELAS} subroutines for goldstino interactions based
on the effective Lagrangian below, and implement them 
into~{\tt MadGraph/MadEvent(MG/ME)v4}~\cite{Stelzer:1994ta,Maltoni:2002qb,Alwall:2007st},
as an alternative to the code for gravitinos~\cite{Hagiwara:2010pi}.%
\footnote{The Fortran code for simulations of the goldstinos/gravitinos
is available at the KEK HELAS/MadGraph/ MadEvent Home Page,
{\tt http://madgraph.kek.jp/KEK/}.}
As we will see later, in the high energy limit, the new code for
goldstinos agrees with the code for gravitinos
with the correction of order $m_{3/2}/\sqrt{s}$
due to the goldstino equivalence theorem.
The new code could be useful especially for collider phenomenology,
where the goldstino limit is good approximation for most of the cases.
Practically, calculations of helicity-summed amplitude squared for goldstino
processes are faster than those for gravitino roughly by a factor of four 
due to the 
number of the helicity states and the simpler structures of the HELAS
subroutines. 
We also note that the goldstino code can be applied to models such as
broken global SUSY and goldstini~\cite{Cheung:2010mc}.

The effective interaction Lagrangian for a goldstino in non-derivative
form is~\cite{Moroi:1995fs,Lee:1998aw,Bolz:2000fu} 
\begin{align}
  {\cal L}_{\rm int}
 =&\frac{i\big(m_{\phi^i_{L/R}}^2-m_{f^i}^2\big)}{\sqrt{3}\,\Mpl\,m_{3/2}}
   \big[\bar{\psi}P_{L}f^i(\phi^i_{L})^* 
  -\bar{f^i}P_{R}\psi\,{\phi}^i_{L} \nn\\
  &\hspace*{23mm}-\bar{\psi}P_{R}f^i(\phi^i_{R})^* 
   +\bar{f^i}P_{L}\psi\,{\phi}^i_{R}\, \big] \nn\\
  &-\frac{m_{\lambda}}{4\sqrt{6}\,\Mpl\,m_{3/2}} 
   \bar{\psi}[\gamma^{\mu},\gamma^{\nu}]
   \lambda^{(\alpha)a}F_{\mu\nu}^{(\alpha)a} \nn\\
  &+\frac{ig_{\alpha}m_{\lambda}}{\sqrt{6}\,\Mpl\,m_{3/2}} 
   \bar{\psi}\gamma_5\lambda^{(\alpha)a}\,
   {\phi^i}^*T^{(\alpha)a}_{ij}\phi^j
\label{L_int}
\end{align}
with the reduced Planck mass 
$\Mpl\equiv M_{\rm Pl}/\sqrt{8\pi} 
 \sim 2.4\times 10^{18}$ GeV and the gravitino mass $m_{3/2}$.
$\psi$ is the Majorana-spinor goldstino field, $f^i$ and $\phi^i$
are spinor and scalar fields in the same chiral supermultiplet, and
$P_{R/L}=\frac{1}{2}(1\pm\gamma_5)$ are the chiral-projection operators.
$T^{(\alpha=3,2,1)a}$ are the $SU(3)_C$ $(a=1,\cdots,8)$, 
$SU(2)_L$ $(a=1,2,3)$ and $U(1)_Y$ generators, 
respectively, and $g_{\alpha=3,2,1}$ are the corresponding gauge
couplings. 
The field-strength tensors for each gauge group are
\begin{align}
 F_{\mu\nu}^{(3)a}&= \partial_{\mu}^{}A_{\nu}^{a}
  -\partial_{\nu}^{}A_{\mu}^{a}-g_3f_3^{abc}A_{\mu}^{b}A_{\nu}^{c}, \\
 F_{\mu\nu}^{(2)a}&= \partial_{\mu}^{}W_{\nu}^{a}
  -\partial_{\nu}^{}W_{\mu}^{a}-g_2f_2^{abc}W_{\mu}^{b}W_{\nu}^{c}, \\
 F_{\mu\nu}^{(1)a}&= \partial_{\mu}^{}B_{\nu}^{}
  -\partial_{\nu}^{}B_{\mu}^{},
\end{align}
and the corresponding gauginos $\lambda^{(\alpha=3,2,1)a}$ are gluinos
($\tilde g^a$), winos ($\tilde W^a$) and bino ($\tilde B$), respectively.

The equivalence of the non-derivative form in \eqref{L_int} and the
derivative form which is obtained by the replacement of the gravitino
field $\psi_{\mu}\sim\sqrt{2/3}\,\partial_{\mu}\psi/m_{3/2}$ in the
gravitino interaction Lagrangian (see, e.g. eq.~(2) in
\cite{Hagiwara:2010pi}) has been proved in \cite{Lee:1998aw}. 
The following features of the interaction Lagrangian \eqref{L_int} are
worth noting:
(i) The $\psi$-$f$-$\phi$-$A_{\mu}$ vertex is absent, while a
new quartic vertex, $\psi$-$\lambda$-$\phi$-$\phi$, exists.
(ii) The couplings are proportional to the mass splitting inside the
supermultiplet, 
$m_{\phi^i}^2-m_{f^i}^2$ and $m_{\lambda}$, 
and inversely proportional to the SUSY-breaking vacuum expectation value
through the gravitino mass 
\begin{align}
 m_{3/2}=\langle F\rangle/\sqrt{3}\,\Mpl.
\end{align}

The paper is organized as follows:
In Sect.~\ref{sec:get} we test our code by making use of the goldstino
equivalence theorem, and 
in Sect.~\ref{sec:sample} we give hadronic total cross sections of
associated gravitino productions with a gluino and a squark as
sample numerical results.
Sect.~\ref{sec:summary} presents our brief summary.
In App.~\ref{sec:helas_new} we give the new {\tt HELAS} subroutines
for goldstinos, and in App.~\ref{sec:mg} we   
describe how to implement the amplitudes into {\tt MG}.

\section{Checking the goldstino-gravitino equivalence}\label{sec:get}

In this section, we check the new {\tt HELAS} subroutines and the
modified {\tt MG} for goldstinos by using the goldstino equivalence
theorem.
As mentioned in introduction, goldstinos appear as the longitudinal
modes of massive gravitinos and their interactions become dominant over
the transverse modes in
high-energy processes.
By using the {\tt MG/ME} package with gravitinos~\cite{Hagiwara:2010pi},
we test an agreement between the goldstino amplitudes
and the gravitino amplitudes in the high-energy limit.

\begin{figure}
 \centering 
 \epsfig{file=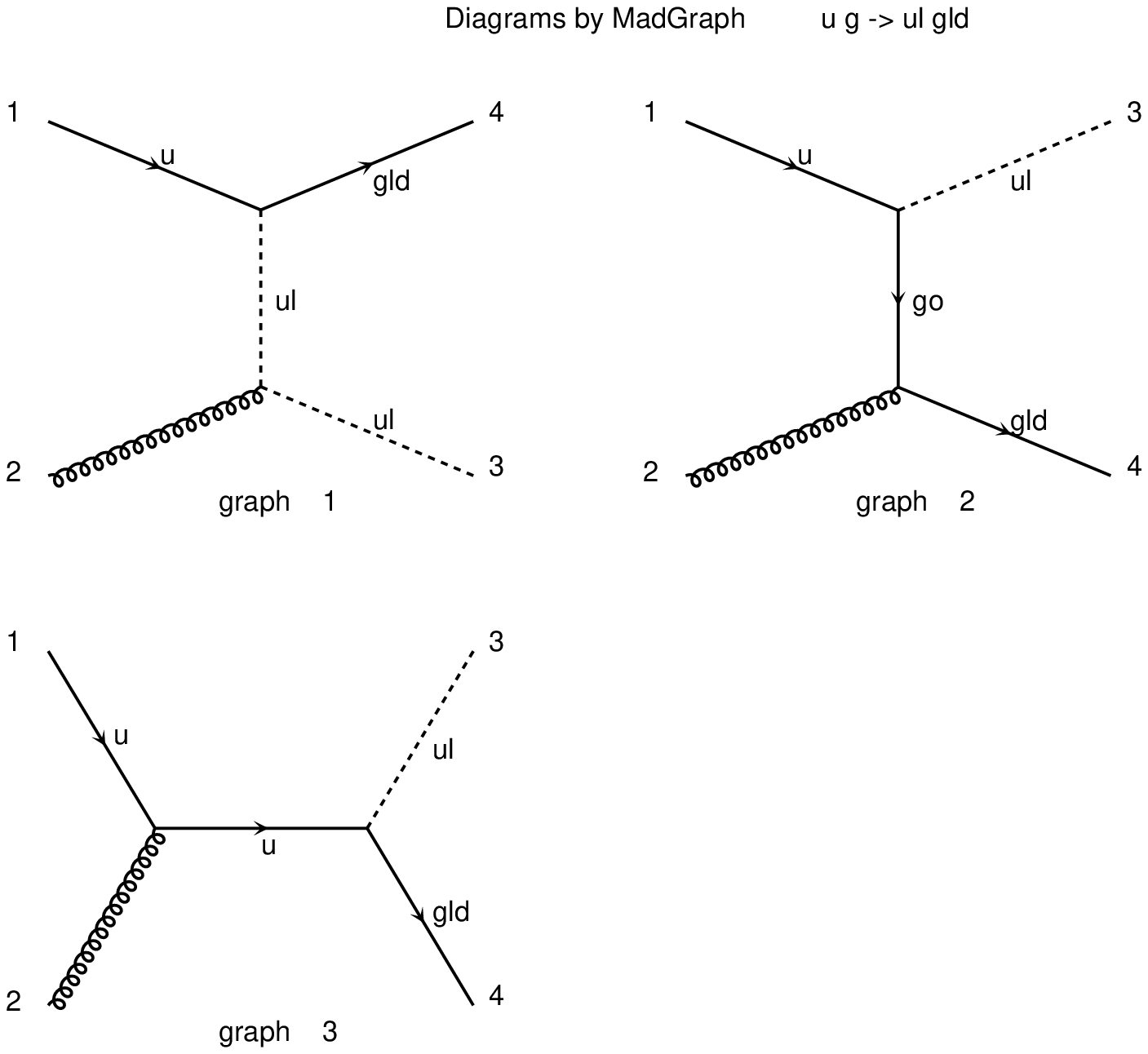,width=1\columnwidth,clip}\\[2mm]
 \epsfig{file=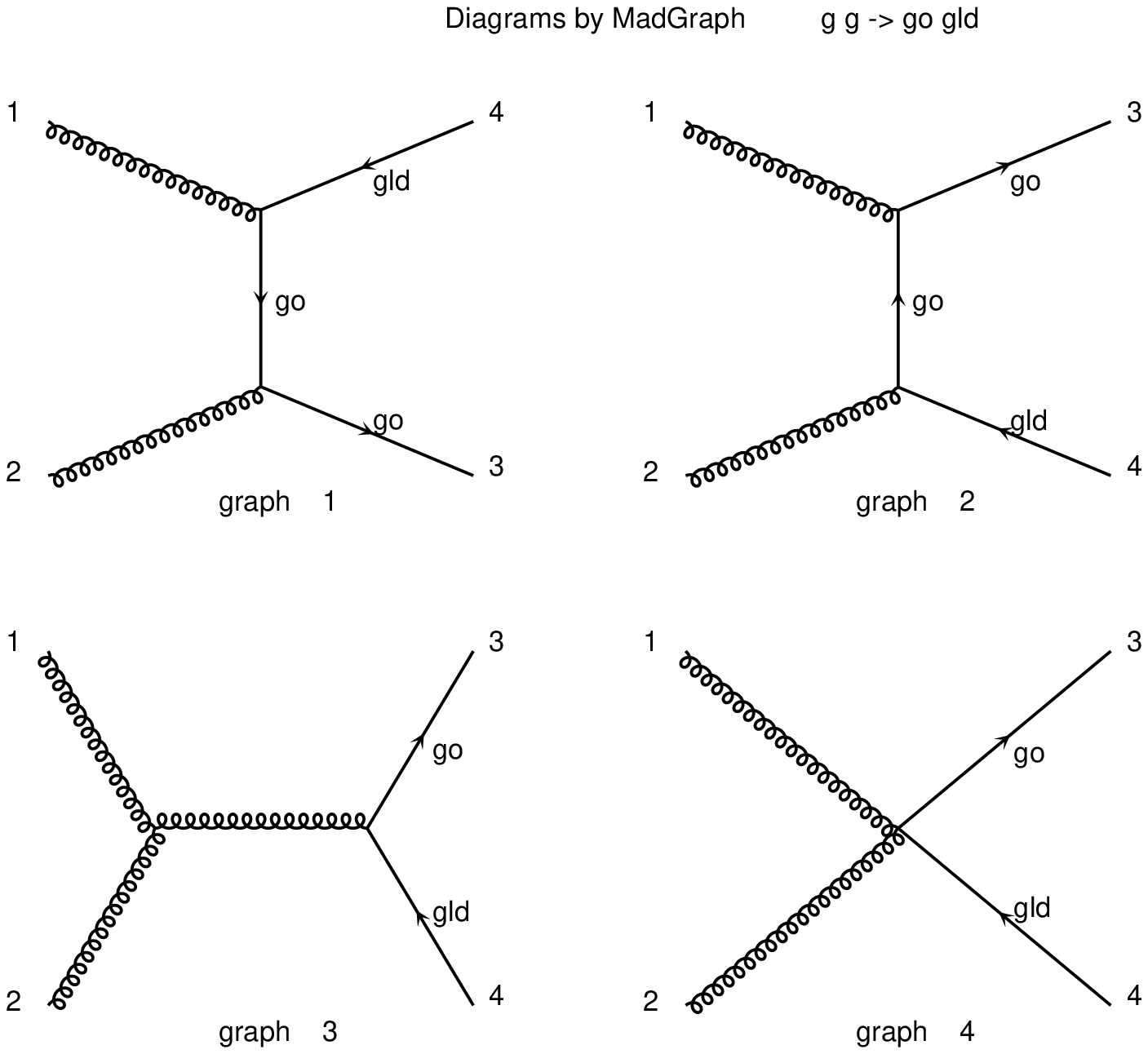,width=1\columnwidth,clip}
 \caption{
  Feynman diagrams for associated goldstino productions with a squark, 
  $qg\to\tilde q\tilde G$ (top), and with a gluino, 
  $gg\to\tilde g\tilde G$
 (bottom), generated by {\tt MadGraph}. 
 {\tt ul}, {\tt go}, and {\tt gld} 
 denote a sup, a gluino, and a goldstino, respectively.
 }
 \label{fig:diagrams}
\end{figure}

We consider the goldstino production processes associated with a squark 
\begin{align}
 q+g\to\tilde q+\tilde G\quad {\rm and}\quad \bar q+g\to\bar\sq+\tilde G,
\end{align}
which involves the $\psi$-$f$-$\phi$ vertex as well as 
$\psi$-$\lambda$-$A_{\mu}$, and with a gluino for the $gg$ initial state
\begin{align}
 g+g\to\tilde g+\tilde G,
\end{align}
which involves the $\psi$-$\lambda$-$A_{\mu}$ and 
$\psi$-$\lambda$-$A_{\mu}$-$A_{\nu}$ vertices.
The Feynman diagrams shown in Fig.~\ref{fig:diagrams} and the
corresponding helicity amplitudes are generated automatically by the
modified {\tt MG}. The details of the {\tt HELAS} subroutines for
goldstinos and those implementation to {\tt MG} are presented in
appendices. 
We note that, in the gravitino case, a diagram with the quartic
$\psi_{\mu}$-$f$-$\phi$-$A_{\nu}$ vertex exists for the squark-gravitino
production. 

\begin{figure}
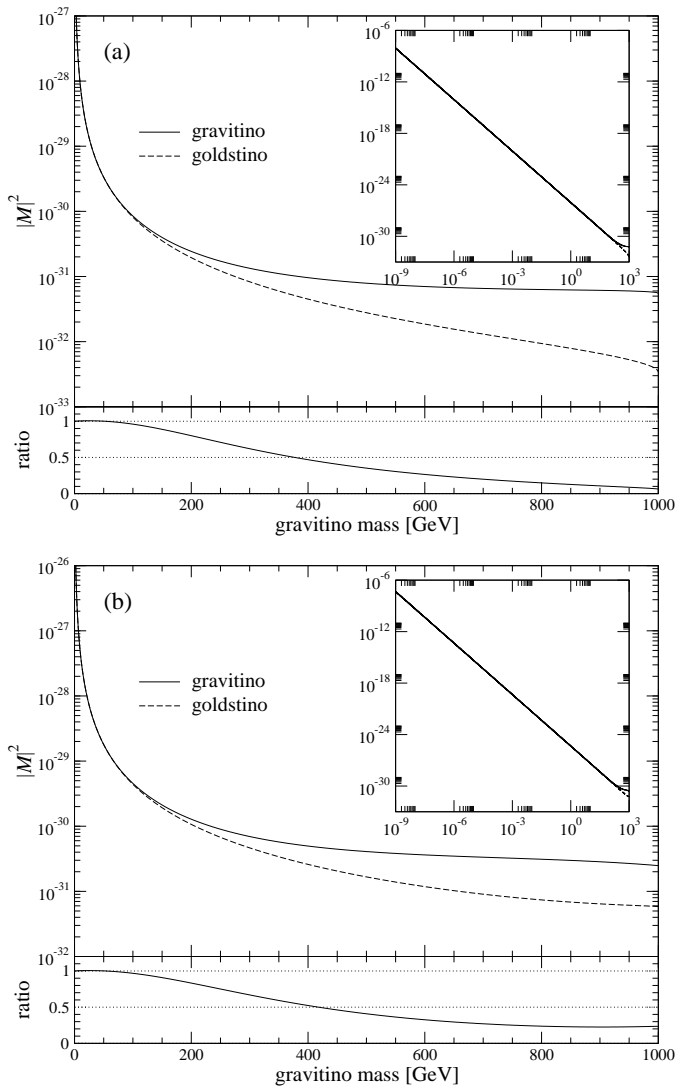

 \centering 
 \epsfig{file=get_qg.eps,width=1\columnwidth,clip}\\[2mm]
 \epsfig{file=get_gg.eps,width=1\columnwidth,clip}
 \caption{
  Squared matrix elements of gravitino (solid) and goldstino (dashed)
  productions associated with a squark, $qg\to\tilde q\tilde G$ (a), and
  with a gluino for the $gg$ initial state,
  $gg\to\tilde g\tilde G$ (b), at $\sqrt{\hat s}=2$ TeV and
 $\cos\hat\theta=0.5$ as a function of the
  gravitino mass, where the 
  squark and gluino masses are fixed at 1 TeV.
  The ratios of the squared matrix elements are also shown. 
  }
 \label{fig:get}
\end{figure}

Figure~\ref{fig:get} shows squared matrix elements of the above two
processes,  
$qg\to\tilde q\tilde G$ (a) and
  $gg\to\tilde g\tilde G$ (b), at the partonic CM energy
$\sqrt{\hat s}=2$ TeV and at the partonic scattering angle
$\cos\hat\theta=0.5$ as a function of the gravitino mass ($m_{3/2}$). 
The squark and gluino masses are fixed at 1 TeV.
In the region of the small gravitino mass, $m_{3/2}/\sqrt{\hat s}< 0.05$,
or in the high energy region $\sqrt{\hat s}\gg m_{3/2}$, the ratio of
the squared matrix elements, 
$|M_{\rm goldstino}|^2/|M_{\rm gravitino}|^2$, is nearly unity, that is,
both amplitudes agree well each other. 
We note that the squared matrix elements for the associated
gravitino productions are proportional to $m_{3/2}^{-2}$ as clearly seen
in the log-log plot in Fig.~\ref{fig:get}.
Associated productions of gravitino and gluino for the $q\bar q$ initial
state
\begin{align}
 q+\bar q\to\tilde g+\tilde G,
\end{align}
which involves the $\psi$-$f$-$\phi$ and  
$\psi$-$\lambda$-$A_{\mu}$ vertices, can be also tested, 
as well as the above processes with an extra parton and crossed processes.

Before turning to sample results, we also note that, 
in addition to the goldstino-gravitino equivalence test,
the code 
was checked carefully by 
comparing with the analytical squared matrix elements of
 eqs.~(3), (5) and (7) in ref.~\cite{Klasen:2006kb} for
the above three partonic processes. 
The hadronic total cross sections at the SUSY benchmark points SPS7 and
SPS8 in figs.~8 and 11 
in ref.~\cite{Klasen:2006kb} can also be reproduced with the help of
{\tt ME}~\cite{Maltoni:2002qb,Alwall:2007st}.      
The test by using the gauge invariance of the amplitudes is also
mentioned in App.~\ref{sec:check}.

\section{Sample results}\label{sec:sample}

In this section, we present some sample numerical results, 
using the new {\tt HELAS} subroutines, 
which are presented in Appendix~\ref{sec:helas_new}, 
and the modified {\tt MG}, which is described in Appendix~\ref{sec:mg}. 

In gauge-mediated SUSY breaking scenarios, the gravitino
is often the lightest supersymmetric particle (LSP), and its
phenomenology depends on what  
the next-to-lightest supersymmetric particle (NLSP) is. 
While the lightest neutralino and the lighter stau are often the NLSP in
minimal models of gauge mediation, gluinos can also be the NLSP, e.g.,
in split SUSY models and general gauge mediation models; see review
papers~\cite{Giudice:1998bp,Morrissey:2009tf} and references therein. 

If gluinos are the NLSP and light enough, those productions can be
explored in the early LHC data as well as in the Tevatron, and 
several studies have been performed for 
hadroproductions of a gravitino with a gluino
(or a squark)~\cite{Dicus:1989gg,Kim:1997iwa,Klasen:2006kb}, which lead
to characteristic signals of monojet plus missing energy when a produced
gluino (squark) promptly decays into a gluon (quark) and a LSP gravitino.  
We consider such scenarios for sample results of our code.

\begin{figure}
 \centering 
 \epsfig{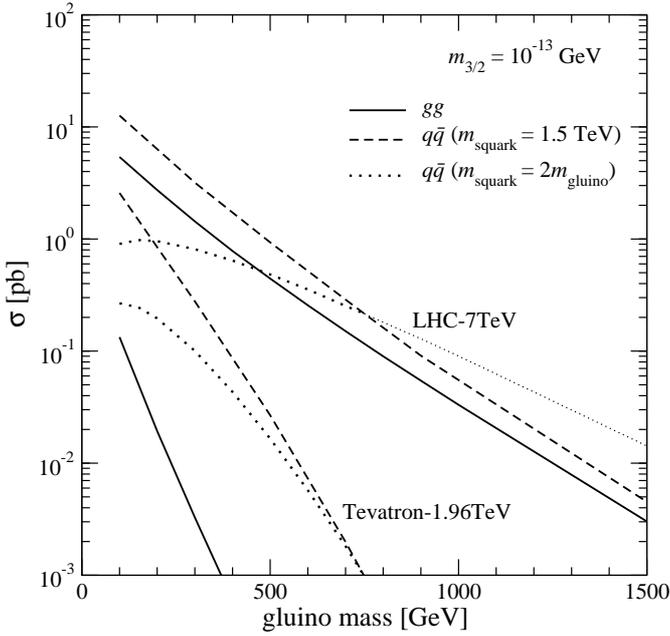}
 \caption{
 Total cross sections of each subprocess of associated gravitino
 productions with a gluino, $p\bar p/pp\to\tilde g\tilde G$, at the
 Tevatron-1.96TeV/LHC-7TeV for $m_{3/2}=10^{-13}$ GeV as a
 function of the gluino mass. 
 The squark masses are fixed at 1.5 TeV (dashed) and $2m_{\tilde g}$ 
 (dotted) for the $q\bar q$ subprocesses,
 where the cross section in the $\Gamma_{\sq\to q\gld}>m_{\sq}/2$ region
 is shown with a thin dotted line.
 }
\label{fig:xsec}
\end{figure}

\begin{figure}
 \centering 
 \epsfig{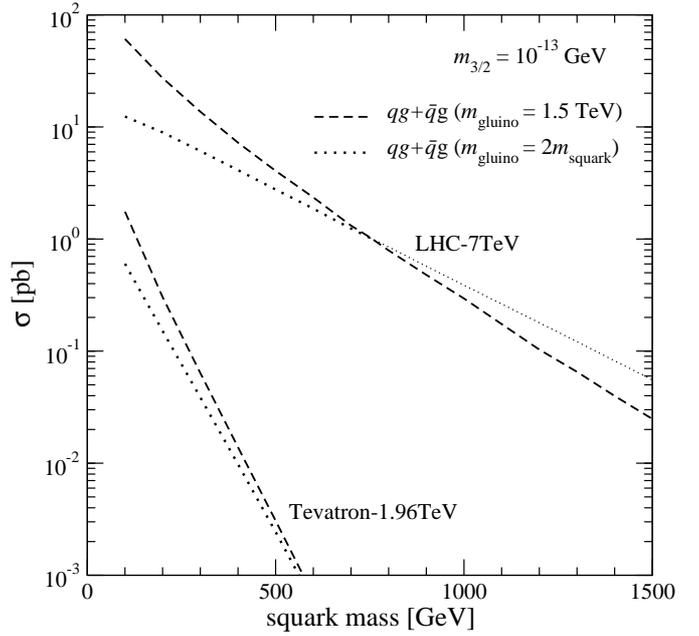}
 \caption{
 Total cross sections of associated gravitino productions with a squark,
 $p\bar p/pp\to\tilde q\tilde G$, at the
 Tevatron-1.96TeV/LHC-7TeV for $m_{3/2}=10^{-13}$ GeV as a
 function of the squark mass. 
 The gluino masses are fixed at 1.5 TeV (dashed) and $2m_{\tilde q}$ 
 (dotted), where the cross section in the $\Gamma_{\go\to
 g\gld}>m_{\go}/2$ region is shown with a thin dotted line. 
 }
\label{fig:xsec_sq}
\end{figure}

Figure~\ref{fig:xsec} presents total cross sections of each subprocess
of associated gravitino productions with a gluino, 
\begin{align}
 p\bar p/pp\to\tilde g\tilde G,
\end{align}
at the Tevatron-1.96TeV/LHC-7TeV for the gravitino mass
$m_{3/2}=10^{-13}$ GeV as a function of the gluino mass.
The masses of the left-handed and right-handed squarks, which appear in
the $t$- and $u$-channel propagators, are fixed at 1.5 TeV (dashed
lines) and $2m_{\tilde g}$ (dotted lines) for the $q\bar q$ subprocesses.
The cross sections of associated productions with a squark,
\begin{align}
 p\bar p/pp\to\tilde q\tilde G\ {\rm and}\ \bar{\tilde q}\tilde G, 
\end{align}
as a function of the squark mass are also shown in
Fig.~\ref{fig:xsec_sq} for reference, where the productions of the 
left-handed and right-handed squarks are summed and their masses 
are taken to be same.
The gluino mass is fixed at 1.5 TeV (dashed lines) and $2m_{\tilde q}$
(dotted lines). 
The CTEQ6L1 parton distribution functions~\cite{Pumplin:2002vw} are
employed, and the renormalization and factorization scales are fixed at
the average mass of the final state particles,
$\mu_R=\mu_F=(m_{\go,\sq}+m_{3/2})/2\sim m_{\go,\sq}/2$.
The cross section in the partial width 
$\Gamma_{\sq(\go)\to q(g)\gld}>m_{\sq(\go)}/2$ region is shown with a thin
dotted line in Fig~\ref{fig:xsec}(\ref{fig:xsec_sq}), which will be
discussed later. 

The major features of the production cross sections are following:
\begin{itemize}
\item The cross sections of all the subprocesses scale with $m_{3/2}^{-2}$,
 that is, lighter gravitinos enhance the monojet signals, which can be 
 interpreted as the direct lower bound for the gravitino mass.%
\footnote{The current lower bound from the
 Tevatron with the integrated luminosity of 87
 pb$^{-1}$ is about 10$^{-14}$ GeV~\cite{Affolder:2000ef}, 
 where all the SUSY particles except gravitino are assumed to be too
 heavy to be produced on-shell~\cite{Brignole:1998me}.}
 Note that, on the other hand, the dijet signals which are produced
 through gluino-pair productions do not depend on the gravitino mass.\\

\item The gluino-gravitino associated productions through the $gg$
 initial state depend only on the gluino mass once the gravitino mass is
 fixed, while the $q\bar q$-initiated production cross section depends
 not only on the gluino mass but also on the $t$- and
 $u$-channel-exchanged squark masses.  
 It should be noted that those contributions are not decoupled in
 the large squark mass limit, and the heavier squark
 exchange increases the cross section since the goldstino-quark-squark
 couplings are proportional to the squark mass squared.
 Therefore, for heavy squark cases, the cross section of the $q\bar q$
 channel can be larger than that of the 
 $gg$ channel even for the $pp$ collider, LHC.\\

\item Similar to the gluino productions through the $q\bar q$ subprocesses,
 the squark-gravitino associated productions, which are only produced
 through the $qg$ subprocesses, involve the $t$-channel gluino
 exchange diagram as well as the squark exchange one (see
 fig.~\ref{fig:diagrams}(top)), and the heavy gluino increases the
 cross section.\\
\end{itemize}

Finally, it is worth noting that the partial width of a gluino (squark)
decay into a gluon (massless quark) and a gravitino is given by 
\begin{align}
  \Gamma_{\go(\sq)\to g(q)\gld}
 =\frac{m^5_{\go(\sq)}}{48\pi\Mpl^2 m^2_{3/2}}.
\end{align}
Therefore, as the gravitino mass becomes small and/or the gluino (squark)
mass becomes large, the partial width rapidly grows. For instance,
in the $m_{3/2}=10^{-13}$ GeV case,
$\Gamma=1.12\times 10^{-3}$ GeV for $m_{\go(\sq)}=100$ GeV, while
$\Gamma=850$ GeV for $m_{\go(\sq)}=1.5$ TeV. 
When the width becomes a significant fraction of the mass, the goldstino
coupling becomes strong and the perturbative calculations are unreliable.
Thin dotted lines in Figs.~\ref{fig:xsec} and \ref{fig:xsec_sq} show
such parameter regions.
Moreover, the decay branching ratios depend on the entire SUSY mass
spectrum; see, e.g., ref.~\cite{Klasen:2006kb}.
Therefore, the comprehensive analyses are necessary for the
collider signatures such as monojet plus missing energy.

\section{Summary}\label{sec:summary}

In this paper, we have added new {\tt HELAS} subroutines to calculate
helicity amplitudes with goldstinos, which appear as the longitudinal
modes of massive gravitinos in high energy processes,
into the {\tt HELAS} library.
They are coded in such a way that arbitrary amplitudes with external
goldstinos can be generated automatically by {\tt MadGraph}, after slight
modifications. All the codes have been tested carefully by making use of 
the goldstino equivalence theorem and the gauge invariance of the
helicity amplitudes. 
Total cross sections of associated 
gravitino productions with a gluino and a squark are also
presented for the Tevatron and the LHC as sample results.

\begin{acknowledgement}{\textit{Acknowledgements}}
We wish to thank Kaoru Hagiwara and Alberto Mariotti for valuable
 discussions and comments
and Junichi Kanzaki for putting our code on the web.
We also would like to thank Tilman Plehn and the members of 
the ITP, Uni.~Heidelberg for their warm hospitality, 
where part of this work has been done.
The work presented here has been in part supported by the Concerted
 Research action 
``Supersymmetric Models and their Signatures at the Large Hadron
 Collider'' 
of the Vrije Universiteit Brussel,
by the IISN ``MadGraph'' convention 4.4511.10,
by the Belgian Federal Science Policy Office through the Interuniversity
 Attraction Pole IAP VI/11, 
and by the Grant-in-Aid for Scientific Research (No. 20340064) from the
 Japan Society for the Promotion of Science.  
Y.T. was also supported in part by Institutional Program for Young
 Researcher Overseas Visits.
\end{acknowledgement}

\appendix
\section{HELAS subroutines for goldstinos}\label{sec:helas_new}

In this appendix, we list the contents of all the new {\tt HELAS}
subroutines that are needed to evaluate processes with an external
goldstino based on the effective Lagrangian of \eqref{L_int}.

In Apps.~\ref{sec:vertex_i} to
\ref{sec:vertex_f}, we explain vertex subroutines listed in
Table~\ref{sublist_new}, which compute interactions among a
goldstino with SM and SUSY particles.
Those subroutines which we do not present in this paper are
a {\tt FFSS} type vertex, the last term in \eqref{L_int}.
This contributes to some phenomenology in cosmology,
e.g., the goldstino emission rate through 
$\tilde q \bar{\tilde q}\to\tilde g\tilde G$ and
$\tilde q \tilde g\to\tilde q\tilde G$~\cite{Lee:1998aw},
but it is mostly irrelevant in collider signatures.
In App.~\ref{sec:check} we briefly mention a test of our new subroutines
by making use of the gauge invariance. 

Before turning to vertex subroutines, we note that 
the existing subroutines for flowing-in ({\tt IXXXXX}) and -out
({\tt OXXXXX}) spinor wavefunctions can be applied for a goldstino
without any modification since it is just a spin-1/2 particle.


\begin{table}
\centering
\begin{tabular}{|c|c|c|c|} \hline
 Vertex & Inputs & Output & Subroutine \\ \hline\hline
 FFS & FFS & Amplitude & {\tt IORSGX}, {\tt IROSGX} \\
     & FS  & F         & {\tt FSORGX}, {\tt FSIRGX} \\ 
     & FF  & S         & {\tt HIORGX}, {\tt HIROGX} \\ \hline
 FFV & FFV & Amplitude & {\tt IORVGX}, {\tt IROVGX} \\
     & FV  & F         & {\tt FVORGX}, {\tt FVIRGX} \\
     & FF  & V         & {\tt JIORGX}, {\tt JIROGX} \\ \hline
 FFVV & FFVV & Amplitude & {\tt IORVVG}, {\tt IROVVG} \\
      & FVV  & F         & {\tt FVVORG}, {\tt FVVIRG} \\
      & FFV  & V         & {\tt JVIORG}, {\tt JVIROG} \\ \hline
\end{tabular}
\caption{List of the new vertex subroutines in {\tt HELAS} system.}
\label{sublist_new}
\end{table}

\subsection{FFS vertex}\label{sec:vertex_i}

The {\tt FFS} vertices involving a goldstino are obtained from the
interaction Lagrangian among a fermion ($f$), a goldstino ($\psi$) and a
scalar boson ($S$):
\begin{align}
 {\cal L}_{\tt FFS}=\bar\psi
  ({\tt GC(1)}P_L+{\tt GC(2)}P_R)f\,S^*+{\rm h.c.}
\label{L_FFS}
\end{align}
with the chiral-projection operator
$P_{R/L}=\frac{1}{2}(1\pm\gamma_5)$. 
The interaction vertices are exactly same as those in the present
{\tt HELAS} 
library, i.e., the Yukawa-type vertex.
Therefore, we simply reuse the existing {\tt IOSXXX}, {\tt FSOXXX} and 
{\tt HIOXXX} subroutines for
{\tt IORSGX}, {\tt FSORGX} and {\tt HIORGX}, respectively, while
{\tt IROSGX}, {\tt FSIRGX} and {\tt HIROGX} need coupling
modification due to the Hermitian conjugate as
\begin{align}
 {\tt GC'(1)}=({\tt GC(2)})^*, \nn\\
 {\tt GC'(2)}=({\tt GC(1)})^*,
\label{hccoupling}
\end{align}
in order to reuse {\tt IOSXXX}, {\tt FSIXXX} and {\tt HIOXXX},
respectively. 
The correspondence between the new {\tt HELAS} subroutines and the
existing ones are shown in Table~\ref{sublist_ref}; see also the
HELAS manual~\cite{Hagiwara:1990dw} for details of these subroutines.

\begin{table}
\centering
\begin{tabular}{|c|c|c|} \hline
 Vertex & New subroutine & Reference subroutine \\ \hline\hline
 FFS & {\tt IORSGX}, {\tt IROSGX} & {\tt IOSXXX} \\
     & {\tt FSORGX} & {\tt FSOXXX} \\
     & {\tt FSIRGX} & {\tt FSIXXX} \\ 
     & {\tt HIORGX}, {\tt HIROGX} & {\tt HIOXXX} \\ \hline
\end{tabular}
\caption{Reference {\tt HELAS} subroutines in the present  {\tt HELAS} 
library~\cite{Hagiwara:1990dw} to the new ones.}
\label{sublist_ref}
\end{table}

{\tt GC(1)} and {\tt GC(2)} are the relevant left- and right-coupling
constants. 
For instance, in the case of the quark-goldstino-squark
interaction without squark mixing, $q$-$\tilde G$-$\tilde q_{\alpha}$,
those couplings read  
\begin{align}
 &\begin{cases}
  {\tt GC(1)}={\tt GFFSL(1)}={\tt GFFS}\\ 
  {\tt GC(2)}={\tt GFFSL(2)}=0
 \end{cases} &&{\rm for}\ \alpha=L, \label{GFFSL}\\
 &\begin{cases}
  {\tt GC(1)}={\tt GFFSR(1)}=0\\ 
  {\tt GC(2)}={\tt GFFSR(2)}=-{\tt GFFS}
 \end{cases} &&{\rm for}\ \alpha=R, \label{GFFSR}
\end{align}
with
\begin{align}
 {\tt GFFS} = im_{\tilde q_{\alpha}}^2/\sqrt{3}\,\Mpl\,m_{3/2}
\label{GFFS}
\end{align}
in the quark massless limit.
We note that, in the {\tt HELAS} convention, the factors of $i$ in the
goldstino coupling constants are necessary to give a correct phase 
between the amplitudes which involve particles in minimal supersymmetric
standard model (MSSM) plus goldstinos.

\subsection{FFV vertex}

The {\tt FFV} vertices involving a goldstino are obtained from the
interaction Lagrangian among a fermion, a goldstino fermion and a vector
boson ($V^{\mu}$): 
\begin{align}
 {\cal L}_{\tt FFV}=-i\bar\psi[\gamma^{\mu},\gamma^{\nu}]
  ({\tt GC(1)}P_L+{\tt GC(2)}P_R)f\,\partial_{\mu}^{}V^*_{\nu}
 +{\rm h.c.}
\end{align}
Although both a goldstino and a gaugino are Majorana particles in most
cases, the Hermitian conjugate term is necessary for {\tt MG};
practically, either the first or second term is used in calculations of
amplitudes with the coupling constant 
\begin{align}
 {\tt GC(1)}={\tt GC(2)}={\tt GFFV}=-i
 m_{f}/2\sqrt{6}\,\Mpl\,m_{3/2}
\label{GFFV}
\end{align}
in units of GeV$^{-1}$, where $m_f$ is the gaugino mass. 

We note that the input and output parameters of the subroutines for the 
{\tt FFV} goldstino vertices are completely same as those of the existing
{\tt FFV}-type subroutines, though the vertex structure computed in the
subroutines is different.

\subsubsection{\tt IORVGX}

This subroutine computes an amplitude of the {\tt FFV} goldstino vertex
from wavefunctions of a flowing-in fermion, a flowing-out 
goldstino and a vector boson, and should be called as
\begin{center}
 {\tt CALL IORVGX(FI,FO,VC,GC , VERTEX)}.
\end{center}
The inputs {\tt FI(6)} and {\tt FO(6)} are complex six-dimensional
arrays which consist of the wavefunction and the four-momentum of the
flowing-{\tt I}n and -{\tt O}ut {\tt F}ermion as
\begin{align*}
 p^{\mu}_{\tt I}&=(\Re e{\tt FI(5)},\Re e{\tt FI(6)},
          \Im m{\tt FI(6)},\Im m{\tt FI(5)}),\\
 p^{\mu}_{\tt O}&=(\Re e{\tt FO(5)},\Re e{\tt FO(6)},
          \Im m{\tt FO(6)},\Im m{\tt FO(5)}), 
\end{align*}
respectively. {\tt VC(6)} is a complex six-dimensional array which contains
the {\tt V}ector boson wavefunction and its momentum as
\begin{align*}
 q^{\mu}=(\Re e{\tt VC(5)},\Re e{\tt VC(6)},
          \Im m{\tt VC(6)},\Im m{\tt VC(5)}).
\end{align*}
The input {\tt GC(2)} is the coupling constant in (\ref{GFFV}). The
output {\tt VERTEX} is a complex number:
\begin{align}
 {\tt VERTEX}=({\tt FO})[\not\!q,\not\!V]
  ({\tt GC(1)}P_{L}+{\tt GC(2)}P_{R})({\tt FI}), 
\end{align}
where we use the notations
\begin{align}
 ({\tt FI})&=\begin{pmatrix}
              {\tt FI(1)}\\{\tt FI(2)}\\{\tt FI(3)}\\{\tt FI(4)}
             \end{pmatrix},\\
 ({\tt FO})&= 
  {\tt (FO(1),FO(2),FO(3),FO(4))},\\
  V^{\mu}&={\tt VC}(\mu+1).
\end{align}

\subsubsection{\tt FVORGX}

This subroutine computes an off-shell fermion wavefunction made
from the interaction of a vector boson and a flowing-out 
goldstino by the {\tt FFV} goldstino vertex, and should be called as
\begin{center}
 {\tt CALL FVORGX(FO,VC,GC,FMASS,FWIDTH , FVORG)},
\end{center}
where the inputs {\tt FMASS} and {\tt FWIDTH} are the mass and the width
of the off-shell fermion, $m_F$ and $\Gamma_F$.
The output {\tt FVORG(6)} is a complex six-dimensional array and gives
the off-shell fermion wavefunction multiplied by the fermion propagator
and its momentum as
\begin{multline}
 {\tt (FVORG)}=({\tt FO})[\not\!q,\not\!V]
  (i{\tt GC(1)}P_{L}+i{\tt GC(2)}P_{R}) \\
 \times\frac{i(\not\!p+m_{F})}{p^2-m_{F}^2+im_{F}\Gamma_{F}},
\end{multline}
and
\begin{align}
 {\tt FVORG(5)}&={\tt FO(5)+VC(5)}, \\
 {\tt FVORG(6)}&={\tt FO(5)+VC(6)},
\end{align}
where we use the notation
\begin{align}
 ({\tt FVORG})=&({\tt FVORG(1),FVORG(2),FVORG(3),FVORG(4)}),
\end{align}
and the momentum $p$ is
\begin{align*}
 p^{\mu}=(&\Re e{\tt FVORG(5)},\Re e{\tt FVORG(6)}, \\
          &\Im m{\tt FVORG(6)},\Im m{\tt FVORG(5)}).
\end{align*}

\subsubsection{\tt JIORGX}

This subroutine computes an off-shell vector current made from
the interaction of a flowing-in fermion and a flowing-out
goldstino by the {\tt FFV} goldstino vertex, and should be called as
\begin{center}
 {\tt CALL JIORGX(FI,FO,GC,VMASS,VWIDTH , JIORG)},
\end{center}
where the inputs {\tt VMASS} and {\tt VWIDTH} are the mass and the width
of the vector boson, $m_V$ and $\Gamma_V$.
The output {\tt JIORG(6)} is a complex six-dimensional array and gives
the off-shell vector current multiplied by the vector boson propagator
and its momentum as
\begin{multline}
 {\tt JIORG(\mu+1)}=\frac{i}{q^2-m_{V}^2+im_{V}\Gamma_{V}}
  \left(-g^{\mu\nu}+\frac{q^{\mu}q^{\nu}}{m_{V}^{2}}\right) \\
  \times ({\tt FO})
  [\not\!q,\gamma_{\nu}]
  (i{\tt GC(1)}P_{L}+i{\tt GC(2)}P_{R})({\tt FI})
\end{multline}
for the massive vector boson, or
\begin{multline}
 {\tt JIORG(\mu+1)}=\frac{-i}{q^2} \\
  \times
 ({\tt FO})[\not\!q,\gamma^{\mu}]
 (i{\tt GC(1)}P_{L}+i{\tt GC(2)}P_{R})({\tt FI})
\end{multline}
for the massless vector boson, and 
\begin{align}
 {\tt JIORG(5)}&={\tt -FI(5)+FO(5)},\\
 {\tt JIORG(6)}&={\tt -FI(6)+FO(6)}.
\end{align}
Here, $q$ is the momentum of the off-shell vector boson,
\begin{align*}
  q^{\mu}=&({\tt \Re eJIORG(5),\Re eJIORG(6),\Im mJIORG(6),\Im mJIORG(5)}).
\end{align*}
Note that we use the unitary gauge for the massive vector boson
propagator and the Feynman gauge for the massless one, according to
the {\tt HELAS} convention~\cite{Hagiwara:1990dw}.

\subsubsection{{\tt IROVGX}}

This is essentially the same subroutine as {\tt IORVGX} but modifies the
input coupling {\tt GC} as in \eqref{hccoupling} since this computes the
Hermitian conjugate vertex of {\tt IORVGX}.

\subsubsection{\tt FVIRGX}

This subroutine computes an off-shell fermion wavefunction made
from the interaction of a vector boson and a flowing-in
goldstino by the {\tt FFV} goldstino vertex, and should be called as
\begin{center}
 {\tt CALL FVIRGX(FI,VC,GC,FMASS,FWIDTH , FVIRG)}.
\end{center}
What we compute here is
\begin{multline}
 {\tt (FVIRG)}=\frac{i(\not\!p+m_{F})}{p^2-m_{F}^2+im_{F}\Gamma_{F}}\\
 \times (i{\tt GC(1)}^*P_{R}+i{\tt GC(2)}^*P_{L})
 [\not\!q,\not\!V]({\tt FI}),
\end{multline}
and
\begin{align}
 {\tt FVIRG(5)}&={\tt FI(5)-VC(5)},\\
 {\tt FVIRG(6)}&={\tt FI(6)-VC(6)},
\end{align}
where we use the notation
\begin{align}
 ({\tt FVIRG})=&\begin{pmatrix}
	      {\tt FVIRG(1)}\\
	      {\tt FVIRG(2)}\\
	      {\tt FVIRG(3)}\\
	      {\tt FVIRG(4)}
	     \end{pmatrix},
\end{align}
and the momentum $p$ is
\begin{align*}
 p^{\mu}=(&\Re e{\tt FVIRG(5)},\Re e{\tt FVIRG(6)}, \\
          &\Im m{\tt FVIRG(6)},\Im m{\tt FVIRG(5)}).
\end{align*}

\subsubsection{{\tt JIROGX}}

This is essentially the same subroutine as {\tt JIORGX} but modifies the
input coupling {\tt GC} as in \eqref{hccoupling} since this computes the
Hermitian conjugate vertex of {\tt JIORGX}.

\subsection{FFVV vertex}\label{sec:vertex_f}

The {\tt FFVV} vertices involving a goldstino are obtained from the
interaction Lagrangian among a fermion, a goldstino and two vector
bosons:  
\begin{multline}
 {\cal L}_{\tt FFVV}=if^{abc}
 \bar\psi[\gamma^{\mu},\gamma^{\nu}]
 ({\tt GC(1)}P_L+{\tt GC(2)}P_R)f^{a}\,V_{\mu}^{b}V_{\nu}^{c}\\
 +{\rm h.c.}
\end{multline}
with the structure constant $f^{abc}$, which can be handled by the 
{\tt MG} automatically.
As in the {\tt FFV} vertex case, we need the Hermitian conjugate term
although a goldstino and a gaugino are Majorana particles in most cases.
The coupling constant {\tt GC} is the product of the {\tt FFV} goldstino
coupling constant and the gauge coupling constant of the involving gauge
boson.
For instance, in the case of the goldstino-gluino-gluon-gluon
interaction, $\gld$-$\go$-$g$-$g$, those couplings are
\begin{align}
 {\tt GC(1)}={\tt GC(2)}={\tt GFFV}*{\tt Gs} =-i
 m_{f} g_s/2\sqrt{6}\,\Mpl\,m_{3/2}
\label{GFFVV}
\end{align}
in units of GeV$^{-1}$, where ${\tt Gs}=g_s$ is the strong coupling
constant.

\subsubsection{\tt IORVVG}

This subroutine computes an amplitude of the {\tt FFVV} goldstino vertex
from a flowing-in fermion, a flowing-out goldstino and two vector
bosons, and should be called as 
\begin{center}
 {\tt CALL IORVVG(FI,FO,VA,VB,GC , VERTEX)}.
\end{center}
What we compute here is
\begin{align}
 {\tt VERTEX}=({\tt FO})[\not\!V^a,\not\!V^b]
  ({\tt GC(1)}P_{L}+{\tt GC(2)}P_{R})({\tt FI}), 
\end{align}
where we use the notations
\begin{align}
 V^{a,\mu}&={\tt VA}(\mu+1),\\
 V^{b,\mu}&={\tt VB}(\mu+1).
\end{align}

\subsubsection{\tt FVVORG}

This subroutine computes an off-shell fermion wavefunction made from the
interaction of two vector bosons and a flowing-out goldstino by the
{\tt FFVV} goldstino vertex, and should be called as
\begin{center}
 {\tt CALL FVVORG(FO,VA,VB,GC,FMASS,FWIDTH\ ,\ FVVORG)}.
\end{center}
What we compute here is
\begin{multline}
 {\tt (FVVORG)}=({\tt FO})
  [\not\!V^a,\not\!V^b](i{\tt GC(1)}P_{L}+i{\tt GC(2)}P_{R}) \\
\times\frac{i(\not\!p+m_{F})}{p^2-m_{F}^2+im_{F}\Gamma_{F}},
\end{multline}
and
\begin{align}
 {\tt FVVORG(5)}&={\tt FO(5)+VA(5)+VB(5)},\\
 {\tt FVVORG(6)}&={\tt FO(6)+VA(6)+VB(6)},
\end{align}
where we use the notation
\begin{align}
 ({\tt FVVORG})=&({\tt FVVORG(1),FVVORG(2),FVVORG(3),FVVORG(4)}),
\end{align}
and the momentum $p$ is
\begin{align*}
 p^{\mu}=(&\Re e{\tt FVVORG(5)},\Re e{\tt FVVORG(6)}, \\
          &\Im m{\tt FVVORG(6)},\Im m{\tt FVVORG(5)}).
\end{align*}

\subsubsection{\tt JVIORG}

This subroutine computes an off-shell vector current made from 
the interaction of a vector boson, a flowing-in fermion and
a flowing-out goldstino by the {\tt FFVV} goldstino vertex,
and should be called as
\begin{center}
 {\tt CALL JVIORG(FI,FO,VC,GC,VMASS,VWIDTH , JVIORG)}.
\end{center}
What we compute here is
\begin{multline}
 {\tt JVIORG(\mu+1)}=\frac{i}{q^2-m_{V}^2+im_{V}\Gamma_{V}}
  \left(-g^{\mu\nu}+\frac{q^{\mu}q^{\nu}}{m_{V}^{2}}\right) \\
 \times({\tt FO})[\gamma_{\nu},\not\!V]
 (i{\tt GC(1)}P_{L}+i{\tt GC(2)}P_{R})({\tt FI})
\end{multline}
for the massive vector boson, or
\begin{multline}
 {\tt JVIORG(\mu+1)}=\frac{-i}{q^2} \\
 \times({\tt FO})[\gamma^{\mu},\not\!V]
  (i{\tt GC(1)}P_{L}+i{\tt GC(2)}P_{R})({\tt FI})
\end{multline}
for the massless vector boson, and
\begin{align}
 {\tt JVIORG(5)}&={\tt -FI(5)+FO(5)+VC(5)},\\
 {\tt JVIORG(6)}&={\tt -FI(6)+FO(6)+VC(6)},
\end{align}
where momentum $q$ is
\begin{align*}
 q^{\mu}=(&\Re e{\tt JVIORG(5)},\Re e{\tt JVIORG(6)}, \\
          &\Im m{\tt JVIORG(6)},\Im m{\tt JVIORG(5)}).
\end{align*}

\subsubsection{{\tt IROVVG}}

This is essentially the same subroutine as {\tt IORVVG} but modifies the
input coupling {\tt GC} as in \eqref{hccoupling} since this computes the
Hermitian conjugate vertex of {\tt IORVVG}.

\subsubsection{\tt FVVIRG}

This subroutine computes an off-shell fermion wavefunction made
from the interaction of two vector bosons and a flowing-in
goldstino by the {\tt FFVV} goldstino vertex, and should be called as
\begin{center}
 {\tt CALL FVVIRG(FI,VA,VB,GC,FMASS,FWIDTH\ ,\ FVVIRG)}.
\end{center}
What we compute here is
\begin{multline}
 {\tt (FVVIRG)}=
  \frac{i(\not\!p+m_{F})}{p^2-m_{F}^2+im_{F}\Gamma_{F}} \\
 \times(i{\tt GC(1)}^*P_{R}+i{\tt GC(2)}^*P_{L})
  [\not\!V^a,\not\!V^b]({\tt FI}),
\end{multline}
and
\begin{align}
 {\tt FVVIRG(5)}&={\tt FI(5)-VA(5)-VB(5)},\\
 {\tt FVVIRG(6)}&={\tt FI(6)-VA(6)-VB(6)},
\end{align}
where we use the notation
\begin{align}
 ({\tt FVVIRG})=&\begin{pmatrix}
	      {\tt FVVIRG(1)}\\
	      {\tt FVVIRG(2)}\\
	      {\tt FVVIRG(3)}\\
	      {\tt FVVIRG(4)}
	     \end{pmatrix},
\end{align}
and the momentum $p$ is
\begin{align*}
 p^{\mu}=(&\Re e{\tt FVVIRG(5)},\Re e{\tt FVVIRG(6)}, \\
          &\Im m{\tt FVVIRG(6)},\Im m{\tt FVVIRG(5)}).
\end{align*}

\subsubsection{{\tt JVIROG}}

This is essentially the same subroutine as {\tt JVIORG} but modifies the
input coupling {\tt GC} as in \eqref{hccoupling} since this computes the
Hermitian conjugate vertex of {\tt JVIORG}.

\subsection{Checking for the new HELAS subroutines}\label{sec:check}

In addition to the goldstino-gravitino equivalence test in
Sect.~\ref{sec:get}, the new {\tt HELAS} subroutines are tested by using
the gauge invariance. Helicity amplitudes involving external gluons are
expressed as 

\begin{align} 
 {\cal M}_{\lambda_g} = 
 \epsilon_\mu(p_g,\lambda_g)\,T^{\mu} 
\end{align}
by extracting one of the gluon wavefunctions,
$\epsilon_{\mu}(p_g,\lambda_g)$, with
a momentum $p_g$ and a helicity $\lambda_g$. 
The identity for the $SU(3)$ gauge invariance
\begin{align}
 {p_g}_{\mu}\,T^{\mu}=0
\end{align}
can be used for checking {\tt HELAS} subroutines.
In particular, we test the invariance of the following processes;
\begin{align*}
 &qg\to\tilde q\tilde G
  &{\rm for}\ &{\tt FFS}\ {\rm and}\ {\tt FFV}\ {\rm subroutines},\\ 
 &gg\to\tilde g\tilde Gg
  &{\rm for}\ &{\tt FFV}\ {\rm and}\ {\tt FFVV}\ {\rm subroutines}.
\end{align*}
We also test the agreement of the helicity-summed amplitude squared
evaluated in arbitrary Lorentz frames.

\section{Implementation of goldstinos into MadGraph}\label{sec:mg}

In this appendix, we describe how we implement 
goldstinos and their interactions into {\tt MG}.

\begin{table}[t]
\centering
\begin{tabular}{ccllllccl}
\hline\hline
\multicolumn{7}{l}{3-point couplings}  & {\tt GC}  & \\
\hline
 FFS && {\tt q} & {\tt gld} & {\tt ql} &&& {\tt GFFSL} & \\
     && {\tt q} & {\tt gld} & {\tt qr} &&& {\tt GFFSR} & \\
 FFV && {\tt go} & {\tt gld} & {\tt g} &&& {\tt GFFV} & \\
\hline\hline
\multicolumn{7}{l}{4-point couplings}        &{\tt GC}   &  \\
\hline
 FFVV &&{\tt go} &{\tt gld} &{\tt g} &{\tt g} &&{\tt GFFVV} &\!\!\!\!\!= {\tt GFFV*Gs} \\
\hline
\end{tabular}
\caption{List of the coupling constants for each goldstino vertex involving  
SUSY QCD particles. All the
 particles and the coupling constants are written in the {\tt MG}
 notation. {\tt gld} stands for a goldstino, 
 {\tt q} represents a light quark, 
 and {\tt ql}/{\tt qr} is a left/right-handed squark.
 {\tt g} and {\tt go} are a gluon and a gluino, respectively.
 {\tt GC} is a coupling
 constant defined in each subroutine in App.~\ref{sec:helas_new}.}
\label{couplist}
\end{table}

First, using the default {\tt mssm} model in 
{\tt MG/ME\,v4}~\cite{Alwall:2007st}, we make
our new model directory, {\tt mssm\_goldstino}, where we define  
a goldstino ({\tt particles.dat}) and its interactions with 
SM and SUSY particles ({\tt interactions.dat} and {\tt couplings.f});
we show the coupling constants for each goldstino vertex involving 
SUSY QCD particles in Table~\ref{couplist} as examples.%
\footnote{It should be noted that the effective
Lagrangian~({\ref{L_int}}) allow to calculate the {\tt FFV} and
{\tt FFVV} interactions only for massless gauge bosons.}
Then we add all the new {\tt HELAS} subroutines
for goldstinos to the {\tt HELAS} library in {\tt MG}.
Although the present {\tt MG} can handle spin-1/2 particles, 
we simply replace spin-3/2 gravitinos in the recent modified
{\tt MG}~\cite{Hagiwara:2010pi} by goldstinos since 
the {\tt FFV} goldstino vertex is non-renormalizable and 
has to be distinguished by the existing renormalizable {\tt FFV} vertex 
and the flowing-in and -out goldstinos require
different subroutines as in the gravitino case.
We note that, in order to achieve the above simple modification, we put 
``{\tt G}'' into the end of names of each gravitino subroutine like
\begin{align*}
 {\tt IORSXX} \to {\tt IORSGX}.
\end{align*}
%


\end{document}